\documentclass[preprint,showpacs,english,aps]{revtex4-1}

\usepackage{babel,rotating,dcolumn}
\usepackage{colordvi,graphicx,color,amsbsy,amsmath,bm,amssymb}
\usepackage{comment}
\begin{document}
\title[]
{Divergence and convergence of inertial particles in high Reynolds number turbulence}
\author{Thibault Oujia$^{1}$}
\email{thibault.oujia@etu.univ-amu.fr}
\author{Keigo Matsuda$^{1,2}$}
\email{k.matsuda@jamstec.go.jp}
\author{Kai Schneider$^{1}$}
\email{kai.schneider@univ-amu.fr}
\affiliation{$^{1}$ Institut de Math\'ematiques de Marseille (I2M), Aix-Marseille Universit\'e, CNRS and Centrale Marseille, 39 rue F. Joliot-Curie, 13453 Marseille Cedex 13, France}
%
\affiliation{$^{2}$ Center for Earth Information Science and Technology, Japan Agency for Marine-Earth Science and Technology (JAMSTEC), Yokohama, Japan}
\date{\today}
\begin{abstract}
 
Inertial particle data from three-dimensional direct numerical simulations of particle-laden homogeneous isotropic turbulence at high Reynolds number are analyzed using Voronoi tessellation of the particle positions, considering different Stokes numbers. A finite-time measure to quantify the divergence of the particle velocity by determining the volume change rate of the Voronoi cells is proposed.
For inertial particles the probability distribution function (PDF) of the divergence deviates from that for fluid particles. 
Joint PDFs of the divergence and the Voronoi volume illustrate that the divergence is most prominent in cluster regions and less pronounced in void regions. 
For larger volumes the results show negative divergence values which represent cluster formation (i.e. particle convergence) and for small volumes the results show positive divergence values which represents cluster destruction/void formation (i.e. particle divergence).
Moreover, when the Stokes number increases the divergence takes larger values, which gives some evidence why fine clusters are less observed for large Stokes numbers. 
Theoretical analyses further show that the divergence for random particles in random flow satisfies a PDF corresponding to the ratio of two independent variables following normal and gamma distributions in one dimension. Extending this model to three dimensions, the predicted PDF agrees reasonably well with Monte-Carlo simulations and DNS data of fluid particles.

\end{abstract}

\maketitle

%

\section{Introduction}

%
Driven by numerous applications of polydispersed multiphase flow, e.g. the rain formation in atmospheric cloud turbulence, or the mist of droplets in the combustion chamber of automobile or aeronautic engines, an abundant number of experimental, numerical and theoretical studies on inertial particles in turbulence can be found in the literature.
For reviews we refer the reader e.g. to \cite{Shaw03, ToBo09,Elgo19}.

Self-organization of the particle density into cluster and void regions
is hereby a typical feature observed in particle-laden turbulent flows and understanding the dynamics is critical for the required mathematical modeling, e.g. accelerated rain formation triggered by clustering. 
The divergence of the particle velocity, which differs due to inertial effects from the divergence-free fluid velocity, plays a crucial role for this clustering mechanism.

In the following we give some overview on related work.
Early results for clustering in particle-laden turbulent flow have been presented in \cite{EaFe94}, and further progress of understanding clustering in homogeneous isotropic turbulence was well summarized in \cite{MBC12}.
The relationship between the divergence of inertial particle velocity and the background flow field was first derived by Robinson \cite{Robi56} and Maxey \cite{Maxe87}. That is, the divergence is proportional to the second invariant of flow velocity gradient tensor for sufficiently small Stokes numbers, which is defined as the ratio of the particle relaxation time $\tau_p$ to the Kolmogorov time $\tau_\eta$. 
This relationship implies that particles tend to concentrate in low vorticity and high strain rate regions in turbulence. This is referred to as the preferential concentration mechanism.
Many theoretical analyses of the preferential concentration have been established following Maxey's approach:
e.g., \cite{EKR96} and \cite{EKLRS02} showed theoretical analyses on the second-order statistics of particle number density using Maxey's formula of the divergence.
Chun et al. \cite{CKRAC05} analytically predicted the radial distribution function for particle number density, which is an important measure of clustering, essentially using an extension of Maxey's approach.
Esmaily-Moghadam and Mani \cite{EsMa16} proposed a first order correction at larger Stokes number for the divergence formula proposed by Robinson and Maxey. 
To understand inertial particle clustering for large Stokes numbers, Vassilicos' group \cite{GoVa06,CGV06,GoVa08,CoVa09} proposed the sweep-stick mechanism, in which particles are swept by large-scale flow motion while sticking to clusters of stagnation points of Lagrangian fluid acceleration. 
This mechanism also explains the multiscale self-similar structure of inertial particle clustering because the zero-acceleration points show self-similar distributions due to the multiscale nature of turbulence. 
Refs. \cite{BIC15} and \cite{AYMY18} reported that self-similar structure of inertial particle clustering is predicted by theoretical analyses for inertial range of turbulence, applying Maxey's formula to a coarse grained flow field at scales where turbulence time scale is sufficiently larger than the particle relaxation time $\tau_p$.
Being apart from the limitation of Maxey's formula, the statistical model for inertial dynamics of small heavy particles were proposed \cite{GuMe16}. In the model, the divergence was obtained from spatial Lyapunov exponents of particle dynamics considering correlation time of particle motion and flow velocity fluctuation \cite{GuMe11,GVM14,GuMe16}.
Interesting findings are that ergodic multiplicative amplification can contribute to clustering significantly under the condition of rapid turbulent velocity fluctuation.

To confirm the reliability of these mechanisms, it is important to quantify the divergence of particle velocity based on direct numerical simulation (DNS) data. 
Blobs of particles were proposed by Bec {\it et al.}~\cite{Bec2007} to define a scale-dependent volume contraction rate using Maxey's formula to determine the divergence of the particle velocity 
for studying coarse grained inertial particle density in the inertial range of turbulence.
Esmaily-Moghadam and Mani \cite{EsMa16} have evaluated the contraction rate (which corresponds to the divergence of particle velocity) using DNS results to verify their theoretical analysis, but they estimated the contraction rate based on the Lagrangian velocity gradient along the trajectory of a single particle. %
In this paper, we aim to compute the divergence directly from the position and velocity of huge number of particles, using the Voronoi tessellation technique.

\medskip

Voronoi tessellation techniques have been first applied to inertial particle clustering in turbulent flow by \cite{MoBC10}. 
Tagawa et al. \cite{TMPCSL12} used the three-dimensional technique to quantify the clustering of inertial particles and bubbles in homogeneous isotropic turbulence obtained by DNS. They calculated the autocorrelation function of Voronoi volume to quantify the Lagrangian decorrelation time scale of clustering.
Voronoi tesslation is further applied to experimental data \cite{OTCMB14,SCAB17,PBC19} and numerical data \cite{DeMo13,BFMC17} to obtain Lagrangian statistics. 
The influence of Stokes and Reynolds number has been analyzed in \cite{SCAB17}. 
Local cluster analysis of small, settling, inertial particles in isotropic turbulence was performed in \cite{MoBr20} using 3d Voronoi tesselation. 
The influence of Reynolds, Froude and Stokes numbers has been assessed and it was shown that
the standard deviation of the Voronoi volumes is strongly dependent on the behavior of void regions compared to the clustering regions. This questions the use of the variance as criterion for cluster detection.

An inherent difficulty for determining the divergence of the particle velocity is its discrete nature, i.e. it is only defined at particle positions.
To this end we propose in the present study a model for quantifying the divergence using  tessellation of the particle positions. The corresponding time change of the  volume is shown to yield a measure of the particle velocity divergence.
The numerical precision of this technique is assessed by applying it first to fluid particles which do not exhibit self-organization into clusters but have a random distribution due to the volume preserving property of fluid velocity. 
Considering then high Reynolds number direct numerical simulation flow data of inertial particles in homogeneous isotropic turbulence we determine the divergence of the particle velocity and analyze its role for structure formation. The influence of the Stokes number is studied and a theoretical prediction of the PDF of the divergence is proposed for the case of random particles.
%


The outline of the manuscript is the following:
In section~\ref{sec:dns} we summarize the DNS data we analyze and in section~\ref{sec:div} the proposed approach to quantify the divergence of the particle velocity is introduced.
Numerical results are presented in section~\ref{sec:numres}.
Finally, conclusions are drawn in section~\ref{sec:concl}.
The convergence of the divergence computation is discussed in the appendix. 

\section{DNS data}
\label{sec:dns}

We analyze particle position and velocity data obtained by three-dimensional DNS of particle-laden homogeneous isotropic turbulence presented in~\cite{MOHKTK14}.
The DNS was performed for a cubic computational domain with side length of $2\pi$.  
The incompressible Navier--Stokes equation was solved using a fourth-order finite-difference scheme. Statistically steady turbulence was obtained by forcing at large-scales, i.e. for $k < 2.5$. 
Discrete particles were tracked in the Lagrangian framework. 
The equation of particle motion is given by 
\begin{equation}
    d_t {\bm v_p}_j = -\tau_p^{-1}({\bm v_p}_j - {\bm u_p}_j) 
    \label{eq:particle},
\end{equation}
where ${\bm v_p}_j$ is the particle velocity vector and ${\bm u_p}_j$ is the fluid velocity vector at particle position ${\bm x_p}_j$. Note that subscript $p$ denotes the quantity at the position of a particle (e.g., ${\bm u_p}_j \equiv {\bm u}({\bm x_p}_j)$), and the subscript $j$ denotes the particle identification number ($j=1,\cdot,N$, where $N$ is the total number of particles). The Stokes drag was assumed for the drag force: The relaxation time $\tau_p$ is independent of the particle Reynolds number.
The reaction of particle motion to fluid flow was neglected.
See \cite{OBT11, MOHKTK14, MaOn19} for details on the computational method.

The number of grid points for the flow field was $N_g^3 = 512^3$. The kinematic viscosity was $\nu=1.10\times10^{-3}$. 
The RMS of velocity fluctuation $u'$, energy dissipation rate $\epsilon$ and Taylor-microscale-based Reynolds number $Re_\lambda$ ($\equiv u' \lambda/\nu$, where $\lambda$ is the Taylor microscale) of obtained turbulence were $u'=1.01$, $\epsilon=0.344$ and $Re_\lambda = 204$. 
These statistics are average over $20 T_0 \le t \le 30 T_0$, where $T_0 \approx L/(2\pi u')$ and $L$ is the domain length.
Note that Matsuda et al. \cite{MOHKTK14} confirmed that $Re_\lambda = 204$ is large enough to be representative of high Reynolds number turbulence at $k\eta > 0.05$, where $\eta \equiv \nu^{3/4} \epsilon^{-1/4}$ is the Kolmogorov scale (i.e., $\eta=7.90\times10^{-3}$).
It should be also noted that the number of grid points was sufficiently large for resolving the turbulent flow so that $k_{max} \eta$ reached 2.03, where $k_{max} \equiv N_g/2$. 
The number of particles $N$ was $1.5 \times 10^7$ and the Stokes number $St$ ($\equiv \tau_p/\tau_\eta$, where $\tau_\eta \equiv \nu^{1/2}\epsilon^{-1/2}$) was $St = 0.05$, 0.1, 0.2, 0.5, 1.0, 2.0 and 5.0.
Particles with different Stokes numbers were tracked in an identical turbulent flow. 
After the turbulent flow had reached a statistically steady state,
particles were randomly seeded satisfying a Poisson distribution.
For this study, we additionally considered data of randomly distributed particles with fluid velocity at the particle positions. These data are analyzed as the fluid particle case; i.e., $St=0$. 
Note that all statistical results are ensemble averaged over 10 snapshots at time $21 T_0 \le t \le 30 T_0$.

\begin{figure}
\centering
\includegraphics[width=0.95\linewidth]{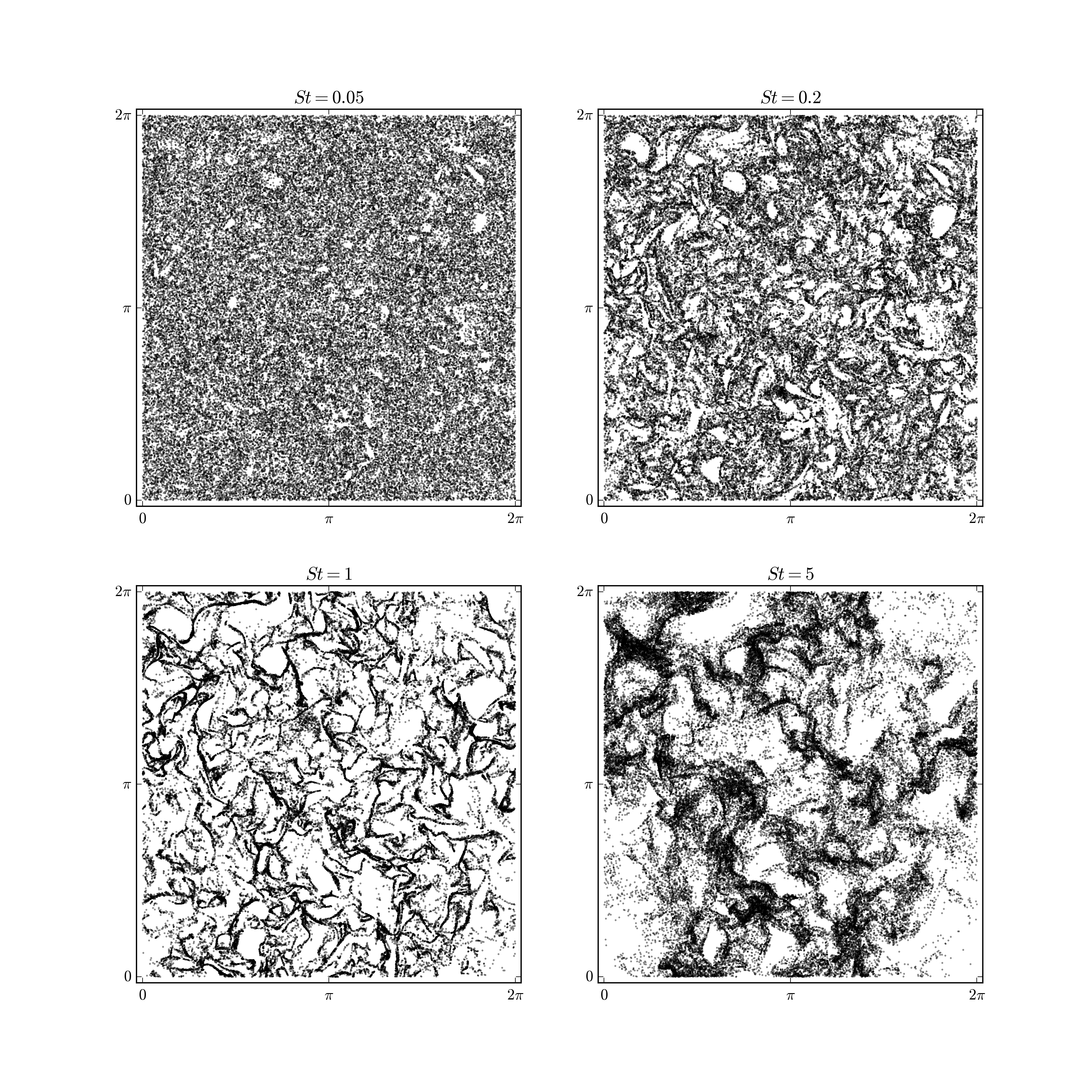}
\caption{Spatial distribution of the particle number density for $St=0.05,0.2,1,5$ at time $t = 24 T_0$ for a slice of thickness $ 4\eta$, where $T_0$ is the representative time scale of DNS. }
\label{fig:Spatial_distribution_of_dropets_St}
\end{figure}

Figure~\ref{fig:Spatial_distribution_of_dropets_St} shows two-dimensional cuts of the particle number density for different Stokes numbers. We can clearly observe void areas for $St=1$. For $St < 1.0$, the void areas are less clear. For $St > 1.0$, they are larger but less pronounced than for $St=1$.

\section{Divergence of the particle velocity}
\label{sec:div}

To understand the dynamics of inertial particles, in particular the clustering, we consider the particle density $n$ in the continuous setting, as shown in many studies (e.g., \cite{Maxe87}). 
It satisfies the conservation equation
which can be written with the Lagrangian derivative, $D_t = \partial_t + {\bm v}\cdot \nabla$, as
%
%
\begin{equation}
    D_t n  = -n \nabla \cdot {\bm v}
    \label{eq:lagrangian_Derivative_n}
\end{equation}
This illustrates that we need to know the divergence of the particle velocity, which is a source term on the r.h.s. in Eq.~(\ref{eq:lagrangian_Derivative_n}).
The problem to determine $\nabla \cdot {\bm v}$ is that we only know the discrete particle distribution, 
and hence we only know the particle velocity at the particle positions, i.e., ${\bm v_p}_j = {\bm v}({\bm x_p}_j)$ but not anywhere else.
To this end Maxey~\cite{Maxe87} proposed a method to obtain the divergence of the particle velocity using the second invariant of the fluid velocity gradient tensor and assuming small $St$.
In this study we aim to compute $\nabla \cdot {\bm v}$ without this assumption.

\subsection{Voronoi tesselation}

The Voronoi tessellation (or diagram) is a technique to construct a decomposition of the space, i.e. the fluid domain, into a finite number of Voronoi cells $C_i$. When a finite number of particles $p_i$ are dispersed in space, a Voronoi cell $C_i$ is defined as a region closer to a particle than other particles. The volume of a Voronoi cell is referred to as Voronoi volume and denoted by $V_{p_i}$.
The  cell $C_i$ can be interpreted as the zone of influence of the particle $p_i$.
The larger the number of particles in a given volume, the smaller the Voronoi volume.
The diagram will allow us to identify particles inside clusters (corresponding to small cells) and particles inside void regions (corresponding to large cells).
A survey on Voronoi diagrams, a classical technique in computational geometry, can be found in Aurenhammer~\cite{Aure91}.

We apply three-dimensional Voronoi tessellation to the DNS data using the Quickhull algorithm provided by the Qhull library in python~\cite{BaDH96}, which has a computational complexity of ${\cal O}(N \log(N))$.

\subsection{Using Voronoi tessellation to compute the divergence}
In the following we propose a method to compute the divergence of the particle velocity ${\cal D}\equiv \nabla \cdot {\bm v}$ in a discrete manner.
Dividing Eq.~(\ref{eq:lagrangian_Derivative_n}) by the particle density $n$, we obtain
\begin{equation}
    {\cal D} = -\frac{1}{n} D_t n
\end{equation}
To calculate the Lagrangian derivative of $n$, we define the local number density $n_p$ as the number density averaged over a Voronoi cell, which is given by the inverse of the Voronoi volume $V_p$; i.e., $n_p = 1/V_p$.
We consider two time instances, $t^k$ and $t^{k+1} = t^k + \Delta t$, where $\Delta t$ is the time step and the superscript $k$ denotes the discrete time index.
The mean number density change in the period of $t^k$ to $t^{k+1}$ is given by $\overline{D_t n}^{\Delta t} = (n_p^{k+1}-n_p^{k})/\Delta t$, where $\overline{n}^{\Delta t} = (n_p^{k+1}+n_p^{k})/2 + {\cal O}(\Delta t)$. Thus, 
we obtain a finite time-discrete divergence of the particle velocity in the period of $t^k$ to $t^{k+1}$,
\begin{equation}
    {\cal D}_p  =  -\frac{2}{\Delta t} \, \frac{n^{k+1}_p-n^{k}_p}{n^{k+1}_p+n^{k}_p} + {\cal O}(\Delta t) 
                =   \frac{2}{\Delta t} \, \frac{V^{k+1}_p-V^{k}_p}{V^{k+1}_p+V^{k}_p} + {\cal O}(\Delta t)
    \label{eq:div_estimator}
\end{equation}
which depends on the choice of $\Delta t$ and $N$. 
This shows that the divergence of the particle velocity can be estimated from subsequent Voronoi volumes, given that the time step is sufficiently small and the number of particles sufficiently large.
To obtain the subsequent Voronoi volumes $V_p^{k+1}$, the particle positions were linearly advanced by ${\bm v_p}$; i.e., ${\bm x_p}^{k+1} = {\bm x_p}^{k} + {\bm v_p}\Delta t$.
The time step was set to $\Delta t = 10^{-3}$, a value which is sufficiently small. The influence of the step size and the number of particles has been checked in Appendix~\ref{appendix1}.

In order to consider the difference with the fluid motion, we also compute the discrete divergence of the fluid velocity 
at Voronoi cells using Eq.~(\ref{eq:div_estimator}). In this case, $V_p^{k+1}$ is obtained from fluid particle positions ${\bm x_p}^{k+1} = {\bm x_p}^{k} + {\bm u_p}\Delta t$ instead.

\begin{figure}[h]
\centering
\includegraphics[width=\linewidth]{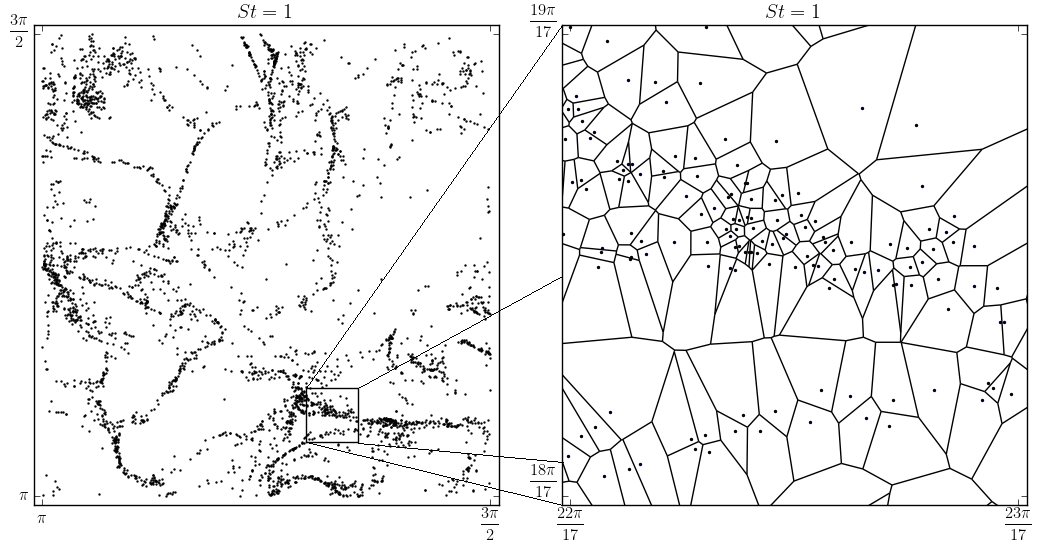}
\caption{Voronoi tesselation generated by particles for $St = 1 $. Two-dimensional particle distribution 
in a slice of thickness $4\eta$ (left), corresponding to a zoom of Fig.~\ref{fig:Spatial_distribution_of_dropets_St} (bottom, left).
A magnified view with Voronoi tessellation (right).
}
\label{fig:Voronoi_diagram}
\end{figure}

Figure~\ref{fig:Voronoi_diagram} shows a two-dimensional particle distribution identical to Figure~\ref{fig:Spatial_distribution_of_dropets_St} ($St=1$) and a magnified view with a corresponding Voronoi diagram. 
We can observe that Voronoi cells of particles in clusters are relatively small, while those corresponding to particles outside clusters are large.

\section{Numerical results}
\label{sec:numres}

In the following, we present numerical results for 
inertial particles in isotropic turbulence considering seven different Stokes numbers.
Fluid particles are likewise analyzed, which allow to understand the statistical properties and to assess the numerical precision of the divergence approximation in Eq.~(\ref{eq:div_estimator}).

\subsection{Inertial particles in turbulence}

First we compute the Voronoi volume for different Stokes numbers and we compare the statistical properties with those obtained for random particles.
\begin{figure}[h]
\centering
\includegraphics[width=0.75\linewidth]{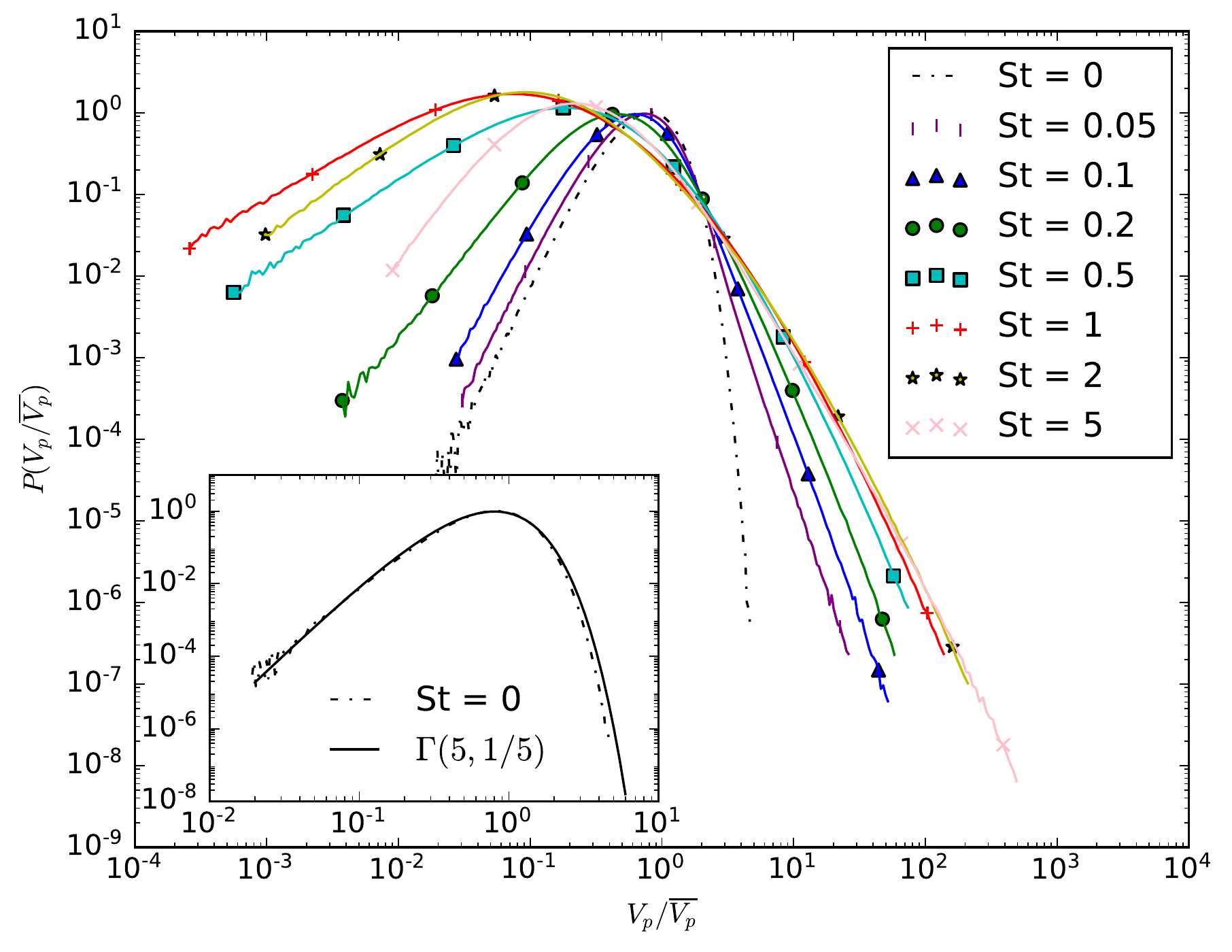}
\caption{PDF of the volume of Voronoi cells in log-log representation normalized by the
mean for different Stokes number and for particles distributed following a Poisson process. The inset compares the PDF for randomly distributed particles (St=0) with the gamma distribution. 
}
\label{fig:StAll_Vor}
\end{figure}
Figure~\ref{fig:StAll_Vor} shows the PDF of the Voronoi volumes $V_p$ obtained from the DNS data, which is normalized by the mean volume $\overline{V_p}=(2\pi)^3/N$. 
For randomly distributed particles, the PDF of the Voronoi volume becomes a gamma distribution~\cite{FeNe07}.
For the 3D case, the PDF of the Voronoi volume is given by $\Gamma(5,1/5)$, where $\Gamma(k,\theta)$ corresponds to the PDF $f_{Vp}(x)=\Gamma(k)^{-1}\theta^{-k}x^{k-1}\exp(-x/\theta)$
and where $k$ and $\theta$ are shape and scale parameters, respectively.
In Fig.~\ref{fig:StAll_Vor} we have confirmed that the PDF for randomly distributed particles in our results agrees well with the gamma distribution of $\Gamma(5,1/5)$.
We can see that, as the Stokes number increases and is getting closer to $1$, the number of small Voronoi cells increases, and then decreases after exceeding St = 1. The number of the large Voronoi cells increases as the Stokes number increases and it stabilizes. 
This $St$ dependence of the PDFs is consistent with the results in previous studies (e.g., \cite{BFMC17}).
The PDF of the Voronoi cell volume is often used to determine ``cluster cells'' and ``void cells''.
In~\cite{OTCMB14} the authors adopted the intersection of the PDF of the Voronoi cell volume and the gamma distribution as thresholds: A Voronoi cell smaller than the smaller threshold (the left intersection) is defined as a cluster cell, and similarly a Voronoi cell larger than the larger threshold (the right intersection) is defined as a void cell. 
In our results, the threshold to determine cluster cells is approximately $V_p/\overline{V_p} \sim 0.5$.

\begin{figure}[ht!]
\centering

\includegraphics[width=0.48\linewidth]{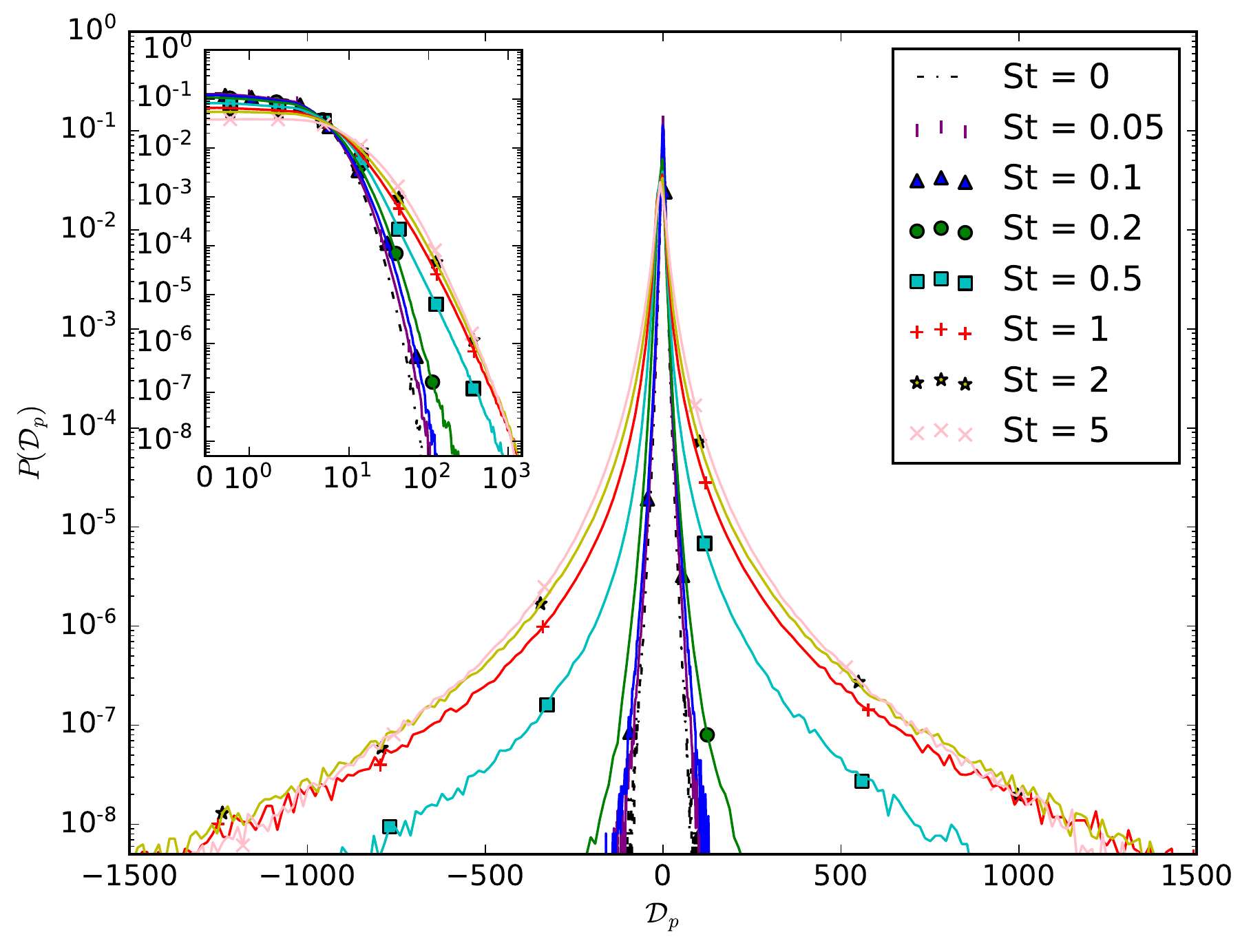}
\includegraphics[width=0.48\linewidth]{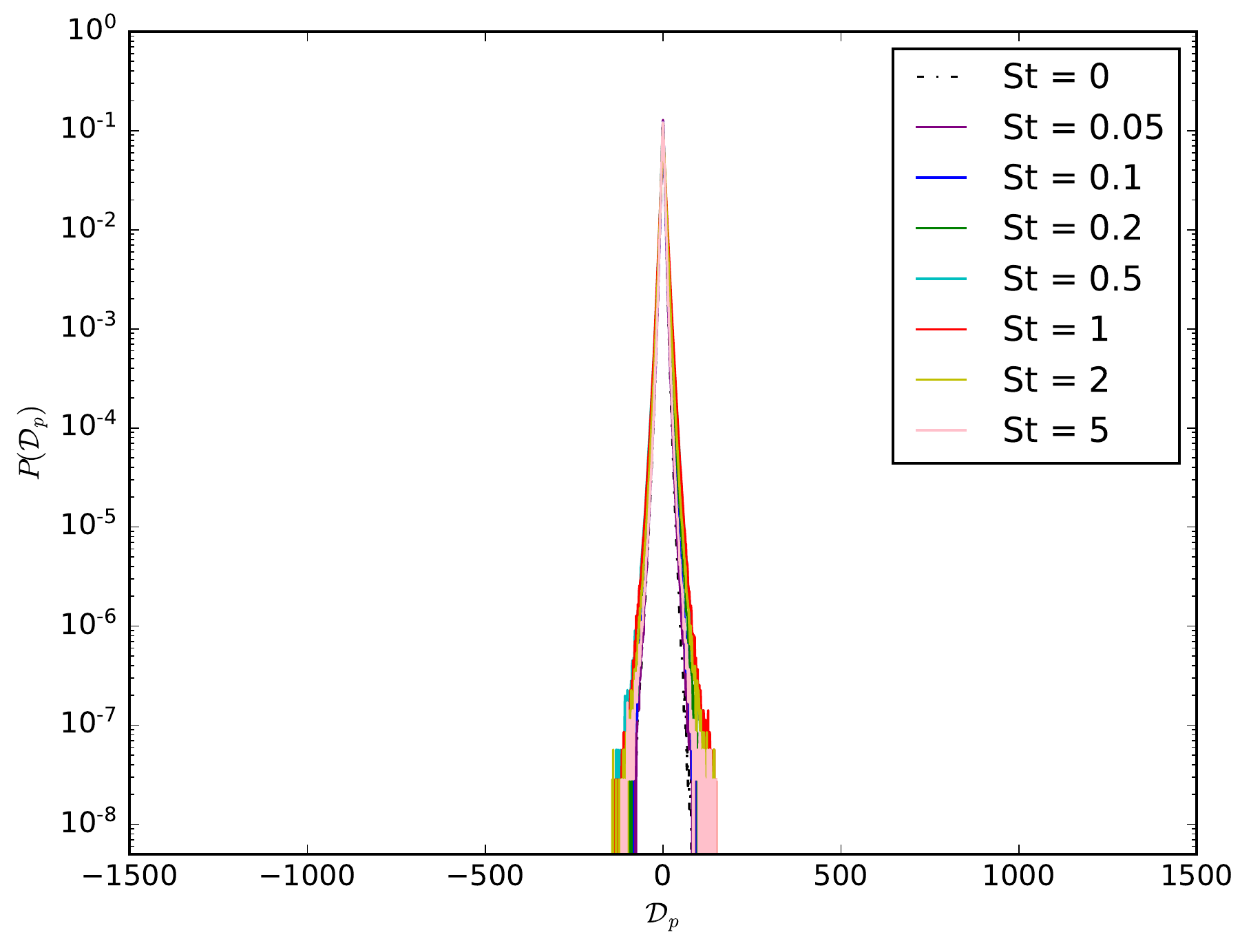} \\

\caption{PDF of divergence for different Stokes numbers: particle velocity (left), fluid velocity (right). The inset (left figure) shows the PDFs for positive divergence values in log-log representation. 
}
\label{fig:Divergence}
\end{figure}
The divergence of the particle velocity is then computed using the time change of the Voronoi volume, given in Eq.~(\ref{eq:div_estimator}).
Figure~\ref{fig:Divergence} shows the PDF of the divergence of the particle velocity (left).  For the particle velocity we observe that the variance increases when Stokes number increases and the tails become heavier. 
For small Stokes numbers ($St \le 0.2$) the variance is strongly reduced.
The PDFs are almost symmetric, centered around 0 with stretched exponential tails.
To understand which part of the divergence of the particle velocity is due to an error of discretization we consider in Fig.~\ref{fig:Divergence} (right) the PDF of the divergence of the fluid velocity. 
Note that for fluid particles in the continuous setting the divergence of the fluid velocity vanishes exactly, while in the discrete setting ${\cal D}_p$  differs from zero, because the deformation of a Voronoi cell is not exactly the same as the deformation of a fluid volume in the continuous setting.
For the divergence of the fluid velocity we find indeed values ranging between $-150$ and $+150$. 
Hence we can deduce that for Stokes numbers less than $0.1$ the divergence is mostly due to an error of discretization, but for larger Stokes numbers, this is a physical effect. 
The numerical precision of the divergence computation has been assessed in appendix~\ref{appendix1}.
%

\begin{figure}[h!]
\centering
\includegraphics[width=0.95\linewidth]{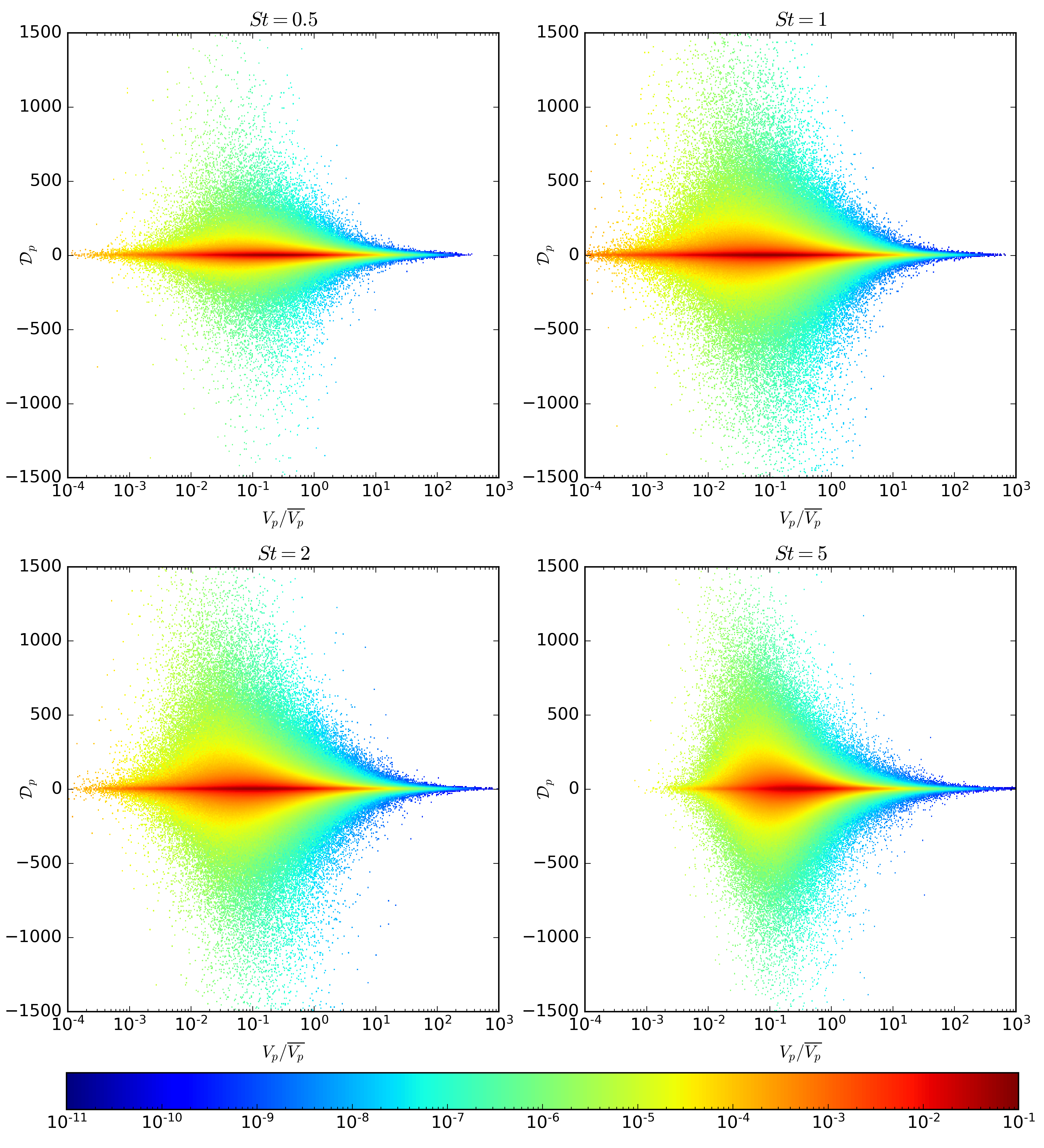}
\caption{Joint PDF of the volume of Voronoi cells in log scale and divergence in linear scale for $St=0.5,~ 1.0,~ 2.0 \text{ and } 5.0$.
}
\label{fig:Joint_PDF_volume_divergence}
\end{figure}

%
To get further insight into the cluster formation, we plot the joint PDF of the divergence ${\cal D}_p$ and the Voronoi volume $V_p$ normalized by its mean. 
Figure~\ref{fig:Joint_PDF_volume_divergence} shows the joint PDF for the cases of $St\ge0.5$ where the variance of ${\cal D}_p$ for inertial particle velocity is larger than that for fluid velocity.
%
For all Stokes numbers, high probability is found along the line of ${\cal D}_p=0$. This indicates that the most of the particles have quite small divergence.
We can find large variance of the divergence that corresponds to the variance shown in Fig.~\ref{fig:Divergence} (left). Such high probability of large positive/negative divergence is observed for cluster cells ($V_p/\overline{V_p} \lesssim 0.5$). 
A possible reason is that the probability of finding particles in cluster regions is higher than in void regions, as shown in Fig.~\ref{fig:StAll_Vor}.
Another possible explanation is that the ``caustics'' \cite{WiMe05} of the particle density, where the particle velocity at a single position is multi-valued, causes the large values of divergence. 
In Fig.~\ref{fig:Joint_PDF_volume_divergence}, it is also observed that large positive divergence values are mostly found at smaller volumes compared to large negative divergence values. This trend becomes more pronounced as the Stokes number increases and suggests asymmetry of the probability of positive and negative $\mathcal D_p$.

\medskip

\begin{figure}[h!]
\centering
\includegraphics[width=0.45\linewidth]{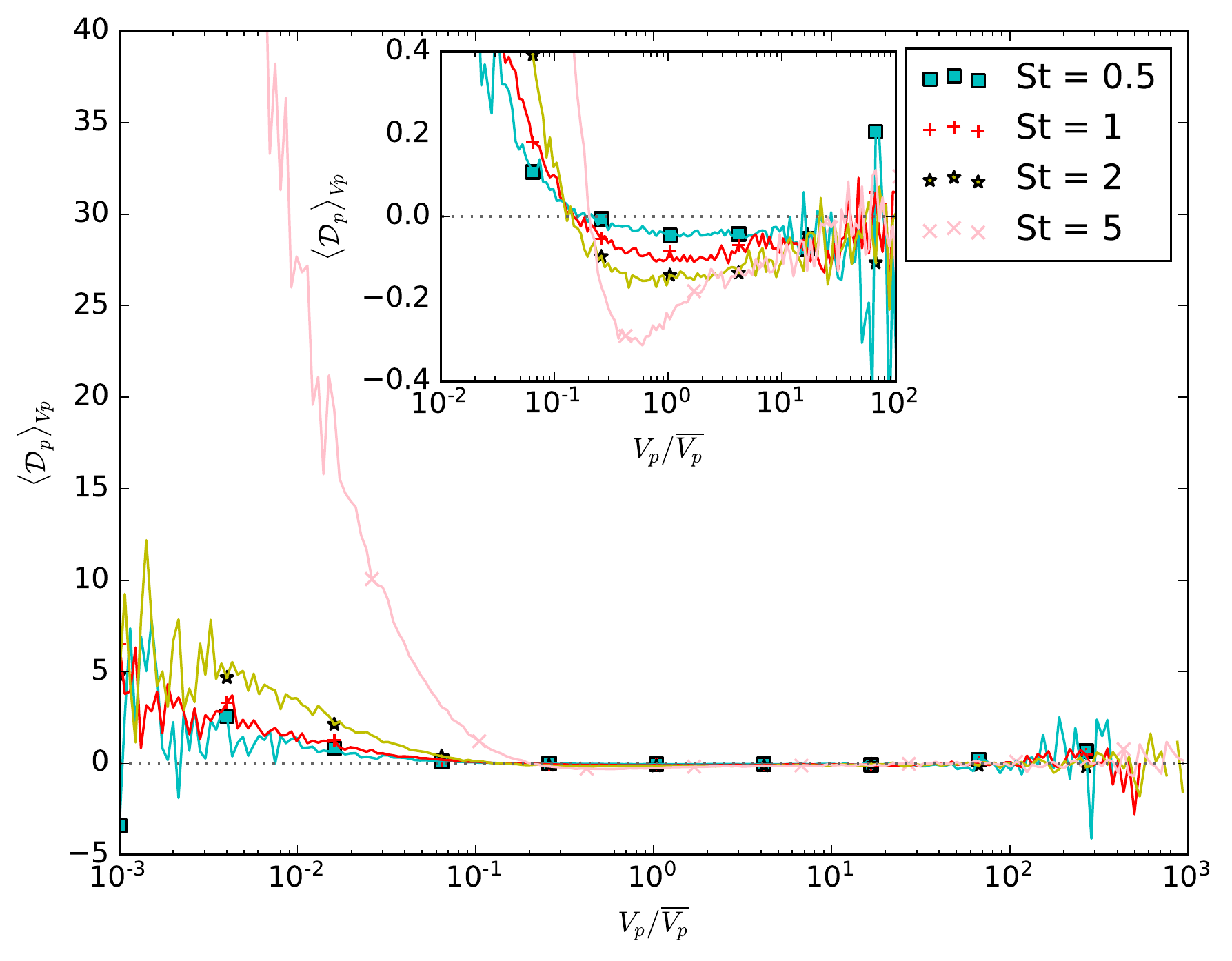}
\includegraphics[width=0.45\linewidth]{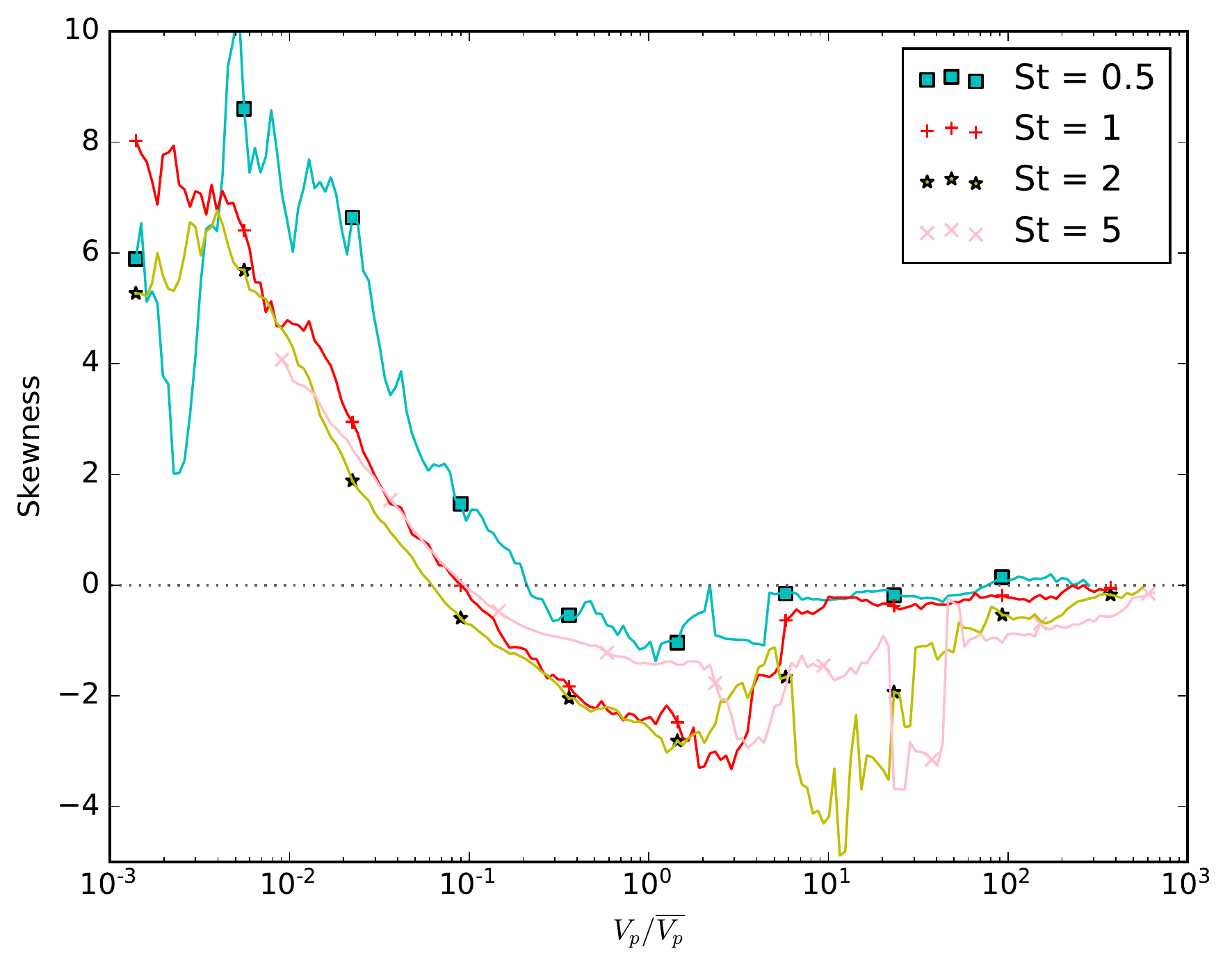}
\caption{Mean value of $\langle {\cal D}_p \rangle_{Vp}$ as a function of the Voronoi volume for different Stokes numbers (left), the inset shows a zoom for large volumes to illustrate negative values.
{
Skewness (right) as a function of the Voronoi volume. Note that for the skewness a moving average filter has been applied. 
} 
}
\label{fig:Joint_PDF_mean_divergence}
\end{figure}

We also compute the mean of the divergence as function of the volume, defined as 
\begin{equation}
    \langle {\cal D}_p \rangle_{Vp} = \frac{1}{P(V_p/\overline{V_p})}\int_{-\infty}^{+\infty} {\cal D}_p \, P({\cal D}_p,V_p/\overline{V_p}) \, d{\cal D}_p
    \label{eq:average_divergence},
\end{equation}
and shown in Fig.~\ref{fig:Joint_PDF_mean_divergence} (left).
When the mean is negative we mostly have convergence of the particles and when the mean is positive we have divergence.
We can see that the conditional average of the divergence $\langle {\cal D}_p \rangle_{Vp}$ is negative for large volumes and turns to positive as the volume becomes smaller. This indicates that the cluster formation is active for large volumes while cluster destruction/void formation is active for small volumes.
\\
Moreover, we observe that when the Stokes number increases the divergence takes larger values for both negative and positive sides. This means that the cluster formation becomes intensified as the Stokes number increases, and the cluster destruction for small volumes is also intensified. This is consistent with the fact that fine clusters are less observed as the Stokes number increases for $St \gtrsim 1$.

To quantify the asymmetry of the joint PDFs in Fig.~\ref{fig:Joint_PDF_volume_divergence} we plot in
Fig.~\ref{fig:Joint_PDF_mean_divergence} (right) the skewness of the divergence as a function of the Voronoi volume. 
We observe a similar trend, positive values for small volumes and negative values for large volumes, for the shown Stokes numbers. 
The negative values of the skewness are more significant than those for $\langle {\cal D}_p \rangle_{Vp}$, and thus we can confirm the asymmetry of the joint PDF at intermediate volumes in the range more clearly, e.g. for $St=1$ from $V_p/\overline{V_p} = 10^{-1}$ to almost $10^{1}$.
For all Stokes numbers, $\langle {\cal D}_p \rangle_{Vp} \approx 0$ and the skewness is close to zero for $V_p/\overline{V_p} > 10^2$. This indicates that the joint PDF is symmetric, which suggests that void formation almost balances void destruction.

\subsection{Randomly distributed particles}

%

To quantify the discretization error of the divergence estimation we consider first a simple toy model in one dimension. We move randomly distributed particles on the line to the right or left.
We assume that the particle velocity satisfies a normal distribution with zero mean and variance $\sigma^2$, i.e., $v_p \sim \mathcal N(0,\sigma^2)$. The volume change $D_t V_p$ is given by the relative velocity of neighbor particles: $D_t V_p = (v_{p,{\rm right}} -v_{p,{\rm left}})/2$, and thus the PDF of the normalized volume change $(D_t V_p)^* \equiv D_t V_p/v_0$ exhibits $\mathcal N(0,\sigma'^2)$, where $\sigma'=\frac{\sigma}{\sqrt{2} v_0}$ and $v_0$ is the representative particle velocity.
The Voronoi volume satisfies a gamma distribution $\Gamma(k,\theta)$ with shape parameter $k=2$ and rate $\theta=1/2$, i.e.
$V_p^* \equiv V_p/\overline{V_p} \sim \Gamma(2,1/2)$.
The PDF of the normalized divergence $\mathcal D_p^* \equiv \mathcal D_p/(\overline{V_p}^{-1} v_0)$ is then given by the product distribution of two independent random variables $X=(D_t V_p)^* \sim \mathcal N(0,\sigma'^2)$ and $Y={V_p^*}^{-1} \sim \Gamma^{-1}(2,2)$, where $\Gamma^{-1}(k,1/\theta)$ is the inverse gamma distribution defined as $f_Y(y) = \frac{\theta^{-k}}{\Gamma(k)}\frac{1}{y^{k+1}}\exp(-1/(\theta y))$. 
%
%
Note that we consider the absolute value of the divergence because the PDF of the divergence is symmetric in this case.
Using the scaled complementary error function ${\rm erfcx}(x) \equiv e^{x^2}\{1 - {\rm erf}(x)\}$, where ${\rm erf}(x)$ is the error function,
we finally obtain the expression for the PDF of $Z=|\mathcal D_p|$: 
\begin{equation}
f_{|\mathcal Dp^*|}(z) = \frac{K_1}{2z^5}\left( \sqrt{2\pi}\sigma'^3(4\sigma'^2+z^2) \text{erfcx}\left( \frac{\sqrt{2}\sigma'}{z}\right) - 4\sigma'^4z \right)
\label{eq:ratio_pdf1d},
\end{equation}
where $K_1$ is a normalization constant.

\begin{figure}[ht!]
\centering
\includegraphics[width=0.48\linewidth]{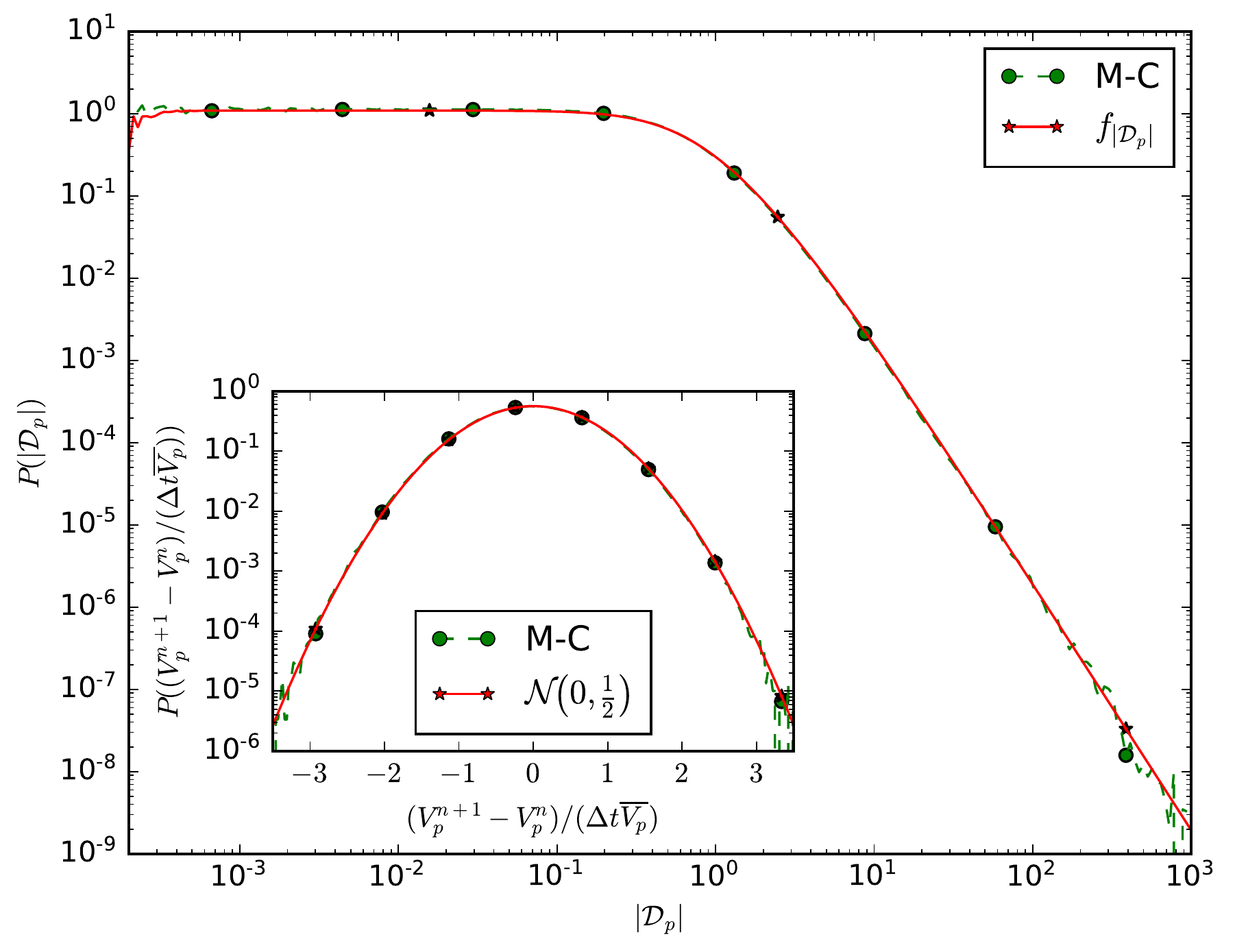}
\includegraphics[width=0.48\linewidth]{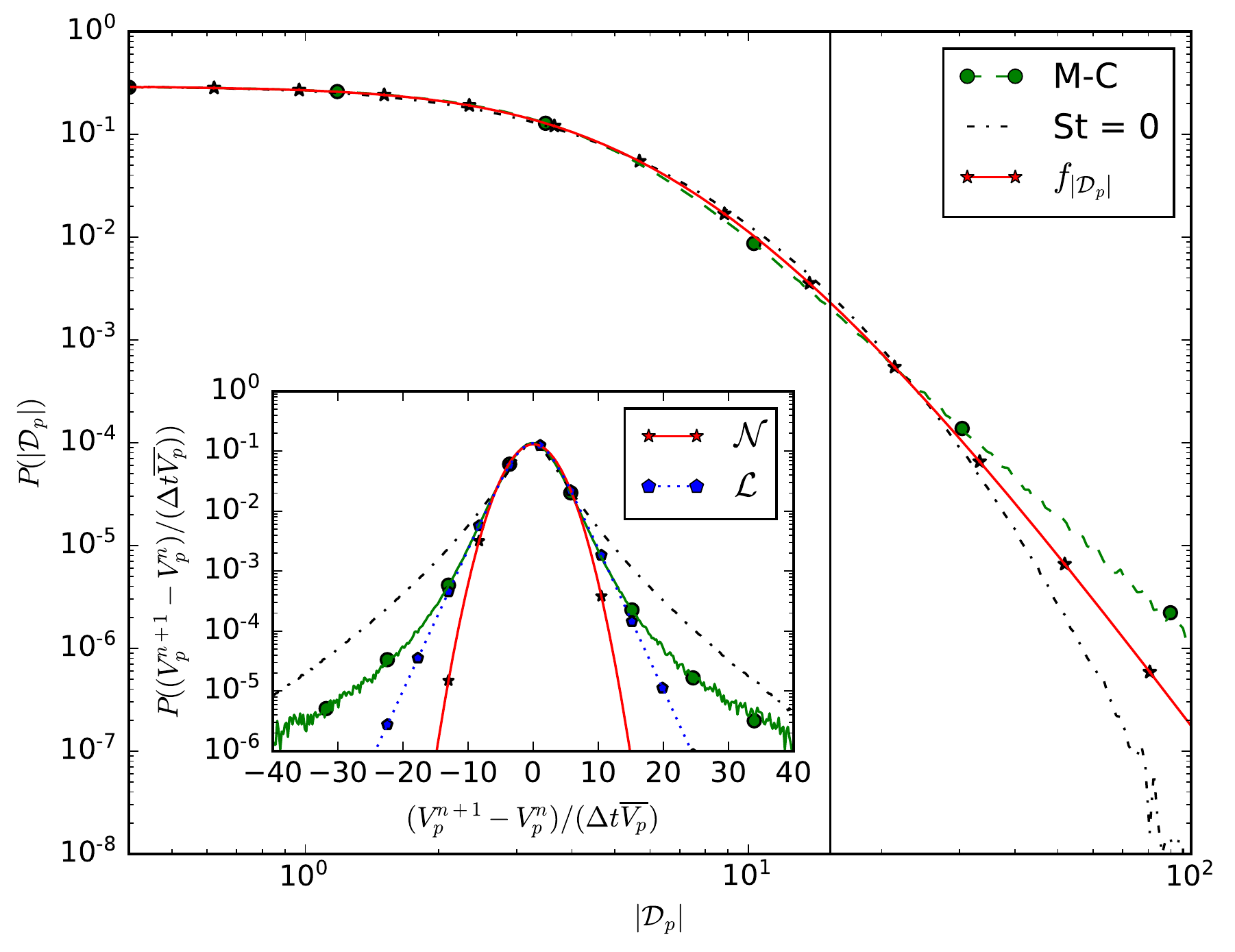}
\caption{PDF of the divergence of the particle velocity using Voronoi analysis in 1D (left) and in 3D (right). 
Shown are the theoretical PDF $f_{|\mathcal D_p|}$,  the Monte--Carlo (M-C) simulation results for Voronoi cells advected by a velocity satisfying a normal law and in the 3D case additionally the DNS results for fluid particles, $St=0$. The vertical line corresponds to 99\% probability. The insets show the corresponding PDFs of the normalized volume changes: Monte-Carlo simulations (green) and $St=0$ (dashed line) in 3D, together with fits for the normal (red) 
(with parameter $\sigma = 3.06$) 
and logistic distribution (blue) (with parameter $s = 1.84$), respectively.
}
\label{fig:mc-sim}
\end{figure}

Figure~\ref{fig:mc-sim} (left) shows the PDF corresponding to the ratio of a normal and a gamma distribution, given by eqn.~(\ref{eq:ratio_pdf1d}), in red and the PDF obtained using Monte-Carlo simulations with, $10^7$ particles. We find perfect agreement between the theory and the numerical simulation.

\medskip

The above model for the divergence of random particles can be extended to three dimensions. 
In 3D the divergence is normalized as $\mathcal D_p^* \equiv \mathcal D_p/(l^{-1} v_0)$, where $l$ is the mean particle distance $l \equiv \overline{V_p}^{1/3}$.
Similarly to the one-dimensional case, we consider the product distribution of two independent random variables $X=|(D_t V_p)^*|$ and $Y={V_p^*}^{-1}$, assuming $(D_t V_p)^* \equiv D_t V_p/(l^{2} v_0) \sim \mathcal{N}(0, \sigma'^2)$ and ${V_p^*}^{-1} \sim \Gamma^{-1}(5, 5)$.
The resulting PDF of $Z=|\mathcal D_p^*|$ becomes
%
\begin{eqnarray}
f_{|\mathcal D_p^*|}(z) 
&=& K_3 \left(\frac{\sigma'}{z}\right)^6 \left[ 8 +9\left(\frac{5\sigma'}{z}\right)^2 +\left(\frac{5\sigma'}{z}\right)^{4} \right. \nonumber \\
& & \left. - \left\{ 15 +10\left(\frac{5\sigma'}{z}\right)^2 +\left(\frac{5\sigma'}{z}\right)^{4}\right\} \sqrt{\pi}\left(\frac{5\sigma'}{\sqrt{2}z}\right)\operatorname{erfcx}\left( \frac{5\sigma'}{\sqrt{2}z} \right) \right]
\label{eq:ration_pdf_3d}
\end{eqnarray}
%
where $K_3$ is the normalization constant for the three-dimensional case. 
Note that evaluating this expression numerically is ill conditioned and some identities and approximations need to be used for stabilization.

Figure~\ref{fig:mc-sim} (right) shows that PDF obtained with the Monte-Carlo simulation perfectly superimposes with the theoretical prediction (Eq.~\ref{eq:ration_pdf_3d}) for values smaller than $\approx 10^1$. For larger values the two curves exhibit some deviation and the observed small deviation is certainly due to the approximation made in~\cite{FeNe07} concerning the choice in the parameters of the gamma distribution.
The insets in Fig.~\ref{fig:mc-sim} show the PDFs of the time change of the Voronoi volume for the 1D and 3D case, respectively. In 1D we observe a perfect superposition of the Monto-Carlo results with the normal distribution, while in 3D this is not the case. A better fit is observed for the logistic distribution. \\
A possible explanation of the heavy tails in the PDF for the 3D case is that the variance of the volume change becomes larger as the Voronoi volume increases. This happens because the larger Voronoi volume tends to have a larger surface area. Large variance of the volume change at large volume then causes heavier tails in the PDF of the volume change than the normal (and logistic) distribution, because the gamma function has an exponential tail.

\section{Conclusions}
\label{sec:concl}

Voronoi tessellation of the particle positions was applied to different homogeneous isotropic turbulent flows at high Reynolds numbers computed by 3D direct numerical simulation.
Random particles and inertial particles with different Stokes numbers were considered.
For analyzing the clustering and void formation of the particles we proposed to compute the volume change rate of the Voronoi cells and we showed that it yields a finite-time measure to quantify the divergence. 
We assessed the numerical precision of this measure by applying it to fluid particles, which are randomly distributed without self-organization due to the volume preserving nature of the fluid velocity.
From this we concluded that sufficiently large Stokes numbers ($St>0.2$) are necessary to obtain physically relevant results.

Considering the joint PDF of the divergence and the Voronoi cell volume illustrates that the divergence is most pronounced in cluster regions of the particles and much reduced in void regions.
We showed that for larger volumes we have negative divergence values which represent cluster formation (i.e. particle convergence) and for small volumes we have positive divergence values which represents cluster destruction/void formation (i.e. particle divergence).

Moreover, we derived the PDF of the divergence of uncorrelated random particles computed with the Voronoi volume change in 1D exactly and showed that it corresponds to 
the ratio of two independent realizations of a normal and a gamma distribution.
Extending this model to 3D we showed that the resulting PDF of the divergence agreed reasonably well with Monte-Carlo simulations and  DNS data of fluid particles.

Finally, our results suggest that when the Stokes number increases the divergence becomes positive for larger volumes, which gives some evidence why for large Stokes numbers fine clusters are less observed for $St \gtrsim 1$.

\section*{Acknowledgements}
%
T.O. acknowledges financial support from I2M.
K.M. thanks financial support from JSPS KAKENHI Grant Number JP17K14598 and I2M, Aix-Marseille University for kind hospitality. 
K.S. thankfully acknowledges financial support from the Agence Nationale de la Recherche (ANR Grant No. 15-CE40-0019), project AIFIT and thanks JAMSTEC for financial support and kind hospitality.
%
Centre de Calcul Intensif d'Aix-Marseille is acknowledged for granting access to its high performance computing resources.
The direct numerical simulation data analysed in this project were obtained using the Earth Simulator supercomputer system of JAMSTEC. 

\clearpage

\appendix
\section{Numerical precision of the divergence computation}
\label{appendix1}
%
\subsection{Reliability of the method}

In the following we assess the reliability of the discrete Voronoi-based divergence computation proposed in Eq.~(\ref{eq:div_estimator}). To this end we consider a stationary periodic velocity field  ${\bm u}$ in two space dimensions, which has no vanishing divergence. We inject $N_p = 10^5$ randomly distributed fluid particles into the $2 \pi$-periodic square domain and advect them for one time step $\Delta t =10^{-2}$ using the explicit Euler scheme. Then we apply Voronoi tesselation and compute the volume change of the Voronoi volumes according to Eq.~(\ref{eq:div_estimator}).  

\begin{figure}[h!]
\centering
\includegraphics[width=\linewidth]{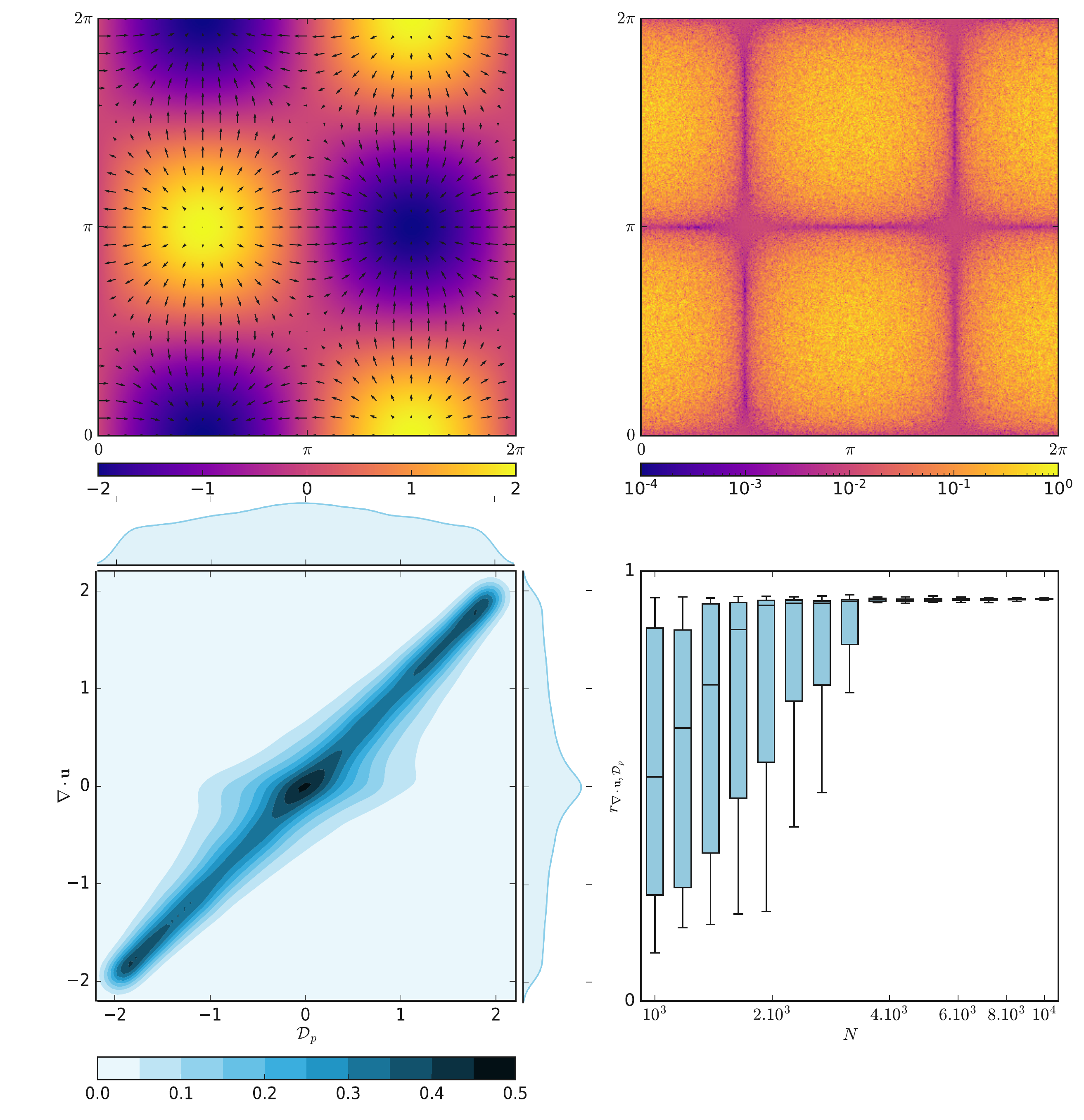}
\caption{
Velocity field ${\bm u}(x, y) = (\cos(x)\cos(y), -\sin(x)\sin(y))$ superimposed with its divergence $\nabla \cdot {\bm u}(x, y) = -2\sin(x)\cos(y)$ (top, left). 
Difference between the exact divergence and the discrete Voronoi divergence in log-scale (top, right).
Joint PDF of the exact divergence and the discrete Voronoi divergence, together with the 1D PDFs of the exact divergence and the Voronoi divergence (bottom, left).
Pearson correlation coefficient between exact and Voronoi divergence as a function of the number of particles $N$ with limit value $r = 0.936$, represented by box plots
(bottom, right).
The horizontal line is the median, the upper and lower box are respectively the first and the third quartile, and top and bottom line is the largest and the lowest data point, respectively, excluding any outliers.
}
\label{fig:divergentvelocityfield}
\end{figure}

Figure~\ref{fig:divergentvelocityfield} (top) superimposes the  velocity field and its divergence (left), and gives the absolute value of the divergence error in log-scale (right).
We observe that the error is most important in strain dominated regions.
Hence we compute the strain rate of the flow, $s_{ij}s_{ij} = 2\cos^2{x}\sin^2{y}$ where $s_{ij}=(\partial_j u_i +\partial_i u_j)/2-\delta_{ij}\partial_k u_k/2$. 
As the correlation between the magnitude of the divergence error and the strain rate is nonlinear we decided to use the Spearman correlation coefficient instead of the Pearson one. We find reasonably well agreement with strong correlation-ship reflected in the value
$r_s = 0.683$.

The joint PDF of the exact divergence and the discrete Voronoi divergence in Fig.~\ref{fig:divergentvelocityfield} (bottom, left) illustrates nicely the strong correlation between the two quantities, as expected. To quantify this correlation we plot in Fig.~\ref{fig:divergentvelocityfield} (bottom, right) the Pearson correlation coefficient for an increasing number of particles, i.e. from $N = 10^3$ to $10^4$ with $15$ values distributed equidistantly in log-scale. 
Shown are box plots indicating the median, and boxes with the first and third quartile to quantify the variability together with min/max values.
After a monotonous increase we observe for $N \ge 4\times10^3$ a saturation at the value of $r = 0.936$, which confirms the strong correlation and also shows that the error does not tend to zero when increasing the number of particles further. However, for increasing number of points the variability is strongly reduced.\\

To summarize, the above discrete divergence results for particles in a given divergent two-dimensional flow are in good agreement with the exact values of the divergence of the carrying flow field.
This shows that the proposed finite-time Voronoi tesselation-based method is well suited and reliable for computing the divergence of the particle velocity.

\subsection{Robustness of the method}

To verify the robustness of the discrete Voronoi-based divergence computations we consider again the previously presented 3D DNS data for $St=0$ and $1$. We check the influence of the time step $\Delta t$ and of the number of particles $N$ on the statistics of the computed discrete divergence of the carrying flow field. Note that in the DNS we have $N = 1.5 \times 10^7$ and $\Delta t =10^{-3}$.
%

\begin{figure}[h]
\centering
\includegraphics[width=0.45\linewidth]{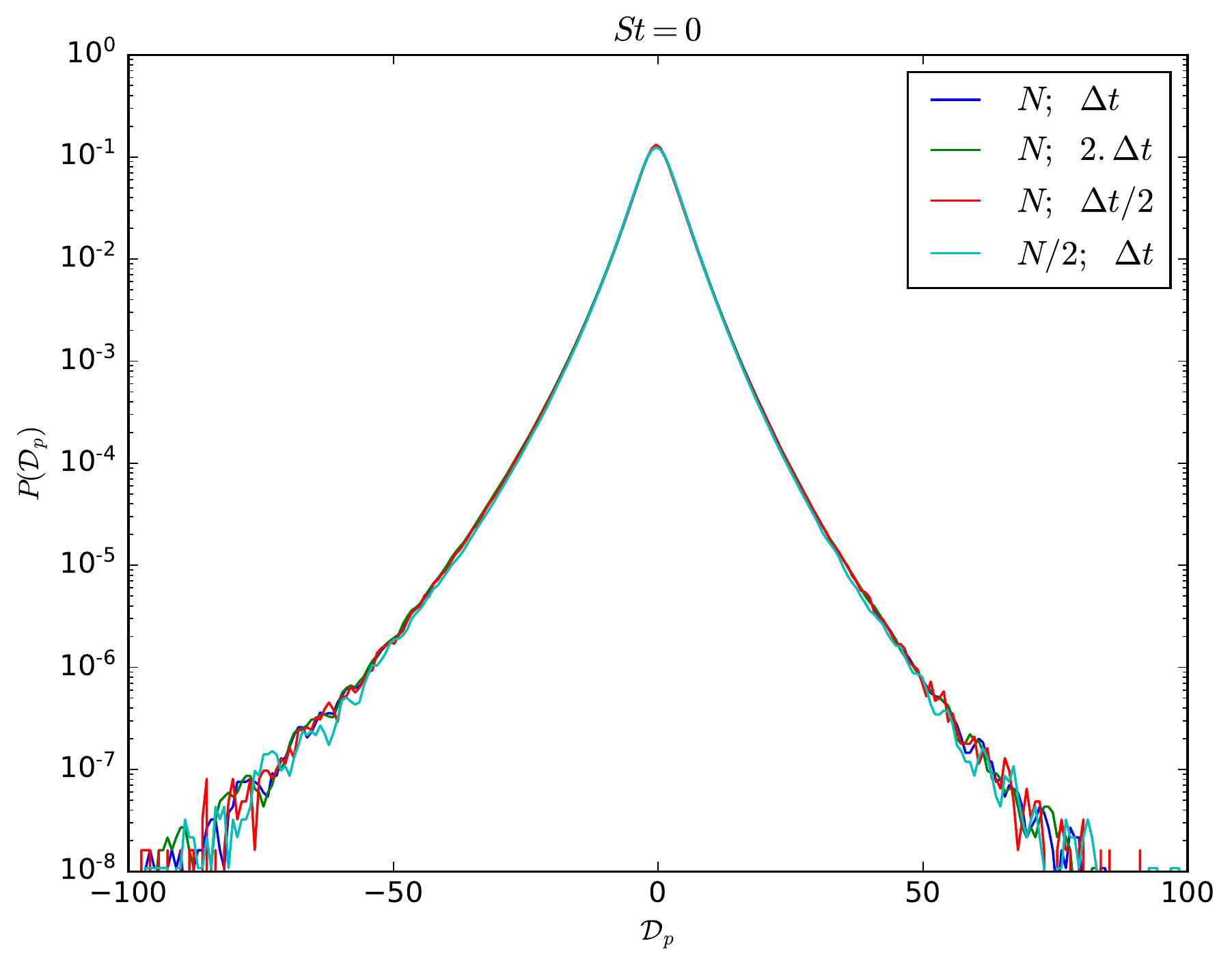}
\includegraphics[width=0.45\linewidth]{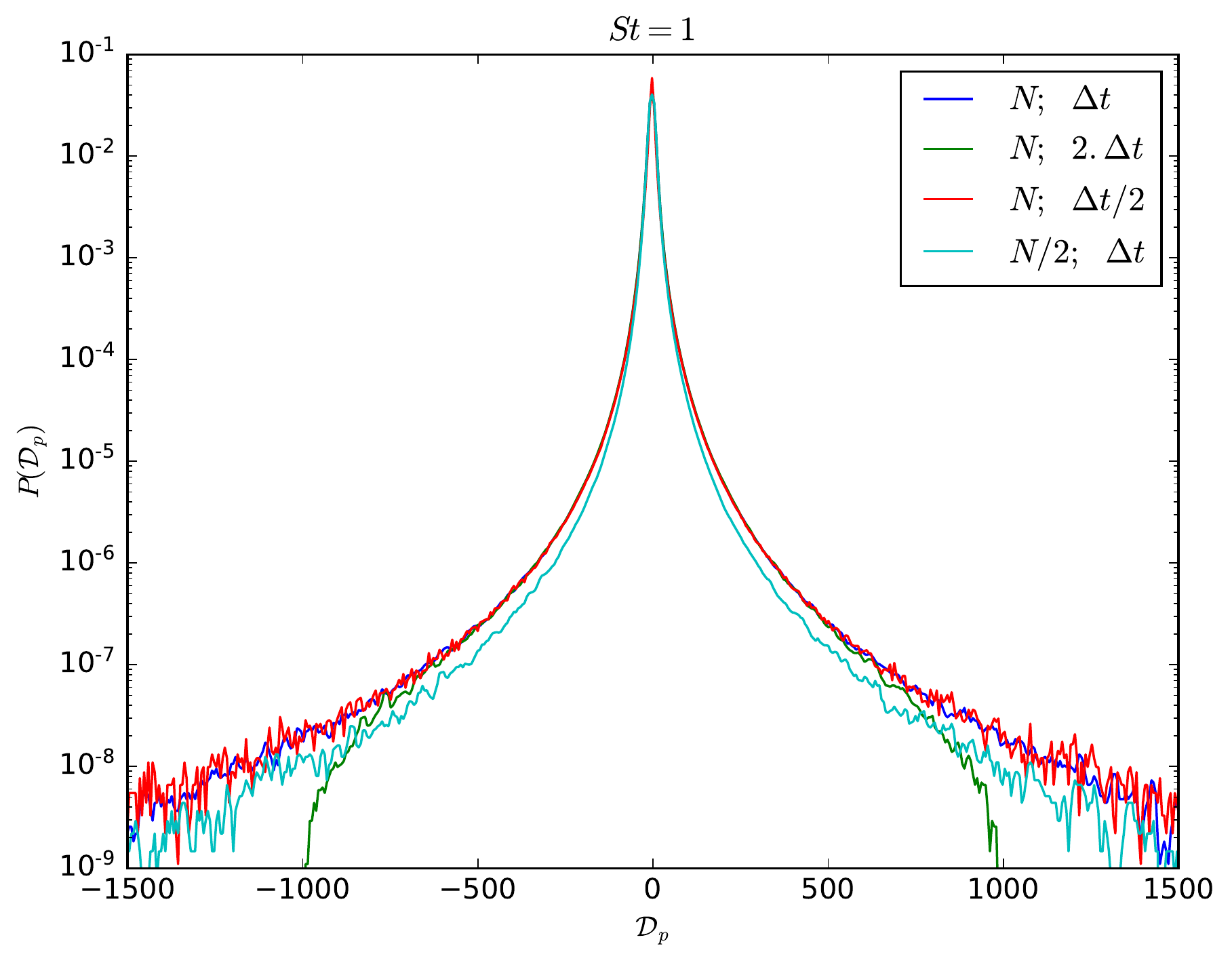}
\caption{PDF of the divergence, ${\cal D}_p$, for $St = 0$ (left) and for $St=1$ (right, where moving average smoothing was applied) for three time steps $\Delta t = 10^{-3}$, $2 \Delta t$ and $ \Delta t / 2$, and two number of particles $N$ and $N/2$. 
}
\label{fig:dt_stability}
\end{figure}

Figure~\ref{fig:dt_stability} shows the PDF of the divergence $\mathcal D_p$ for different values of $\Delta t$ and $N$.
For $St=0$ the PDFs remain almost unchanged and perfectly superimpose when dividing or multiplying the time step by a factor two, or dividing the number of particles by two, which proves the robustness of the statistics for fluid particles.
For $St=1$ the situation changes with the exception of time step reduction, while keeping $N$ fixed. 
Replacing $\Delta t$ by $\Delta t/2$ yields an almost identical distribution. 
This shows that the time step has been chosen sufficiently small.
%
Increasing $\Delta t$ to $2 \Delta t$, while keeping $N$ fixed, some dependence is found for $|\mathcal D_p|>500$ and the tails suddenly decay around $\mathcal D_p = \pm 1000$. 
This can be explained by the CFL condition for the volume change: $|D_t V_{p}| \Delta t/V_{p} < \mathcal O(1)$. 
The divergence is given by $\mathcal D_p = D_t V_{p}/V_{p}$. Thus, 
$\mathcal |D_p| < \mathcal O(1/\Delta t)$ would satisfy the CFL condition. 

%
%
The $N$ dependence of the divergence is due to the $N$ dependence of the mean separation length $l$: The sampling density becomes coarser as $N$ decreases, and then the Voronoi tessellation results loose the information at fine scales. 
This is reflected in lighter tails in the PDF for $N/2$ compared to the one obtained with $N$ particles. However, the extreme values are not significantly modified.
Thus, the $N$ dependence can be considered as the difference in the filtering scale.
The divergence at caustics regions would be sensitive to the filtering scale. 

\medskip

\end{document}